%% file: Paper.tex
\documentclass[final,12pt,numbers]{elsarticle}

\usepackage{amssymb}
\usepackage{amsmath}
\usepackage{algorithm2e}
\usepackage{xcolor} 
\usepackage{booktabs}
\usepackage{lscape}
\usepackage{pgfplotstable}
\usepackage{siunitx}
\usepackage{textcomp}
\usepackage{esvect}
\usepackage{multirow}
\usepackage{adjustbox}
\usepackage{gensymb}
\usepackage{pgfplots}
\usepackage{color, colortbl}
\usepackage{hyperref}
\definecolor{Gray}{gray}{0.9}
\usepgfplotslibrary{fillbetween}
\usetikzlibrary{spy,backgrounds,calc,svg.path}
\pgfplotsset{select coords between index/.style 2 args={
		x filter/.code={
			\ifnum\coordindex<#1\def\pgfmathresult{}\fi
			\ifnum\coordindex>#2\def\pgfmathresult{}\fi
		}
	},
}
\newcommand{\tablesize}{.95} 
\journal{Annals of Nuclear Energy}
\pgfplotsset{compat=1.16}
\begin{document}

\begin{frontmatter}



\title{Gaussian Processes for Surrogate Modeling of Discharged Fuel Nuclide Compositions}


\author[1]{Antonio Figueroa}
\ead{figueroa@aices.rwth-aachen.de}
\author[2]{Malte G\"ottsche}
\ead{goettsche@aices.rwth-aachen.de}

\address{Nuclear Verification and Disarmament Group, RWTH Aachen University \\ Schinkelstra{\ss}e 2a, 52062 Aachen, Germany}



\include{Sections/Abstract}
\begin{highlights}
\item Gaussian Process surrogate models for spent fuel outperform Cubic Spline methods.
\item By nature, GP's estimate prediction errors, a useful quantity for nuclear engineering. 
\item GP's can be useful for cross-sections based methods to estimate output uncertainties.
\end{highlights}


\end{frontmatter}


\input{Sections/Introduction}
\input{Sections/TheoreticalBackground}


\input{Sections/Results}
\input{Sections/Conclusion}

\input{Sections/DataAvailability}
\input{Sections/Acknowledgements}




\bibliography{Sections/Bibliography}
\input{Sections/Appendix}



\end{document}

%% file: Sections/Abstract.tex
\begin{abstract}
     Several applications such as nuclear forensics, nuclear fuel cycle simulations and sensitivity analysis require methods to quickly compute spent fuel nuclide compositions for various irradiation histories. Traditionally, this has been done by interpolating between one-group cross-sections that have been pre-computed from nuclear reactor simulations for a grid of input parameters, using fits such as Cubic Spline. We propose the use of Gaussian Processes (GP) to create surrogate models, which not only provide nuclide compositions, but  also the gradient and estimates of their prediction uncertainty. The former is useful for applications such as forward and inverse optimization problems, the latter for uncertainty quantification applications. For this purpose, we compare GP-based surrogate model performance with Cubic-Spline-based interpolators based on infinite lattice simulations of a CANDU 6 nuclear reactor using the \texttt{SERPENT 2} code, considering burnup and temperature as single dimensional input parameters resulting in a 2D study. Additionally, we compare the performance of uniform grid sampling to quasirandom sampling based on the Sobol sequence. We find that GP-based models perform significantly better in predicting spent fuel compositions than Cubic-Spline-based models on both sampling schemes. Uniform grid sampling performs better than quasirandom sampling in our 2D study, this cannot be generalized to higher dimensions. We have found that GP models require more time to compute a prediction. Although a relatively small time ($\mathcal{O}$ ms), iterative methods used for statistical inference benefit from from as-low-as-possible calculation times. Furthermore, we show that the predicted nuclide uncertainties are reasonably accurate. While in the studied two-dimensional case, grid- and quasirandom sampling provide similar results, quasirandom sampling will be a more effective strategy in higher-dimensional cases.

\end{abstract}


%% file: Sections/Introduction.tex
\section{Introduction}

The computation of spent fuel nuclide compositions is a complex problem which involves  modelling the fuel assembly, reactor geometries and tracking the nuclide evolution as the fuel is irradiated following an operational history indicated by different parameters such as burnup, power, and temperature. This requires the numerical solution of the neutron transport equation through probabilistic - namely Monte Carlo calculations - or deterministic methods \cite{nuchb}, which are computationally expensive. 

For several applications however, it is desirable to have a fast and computationally less intensive method to compute nuclide compositions, especially when many repeated calculations are needed for further analysis. Examples of this are nuclear fuel cycle simulators \cite{cyclus}, nuclear forensics applications \cite{forensics}, and sensitivity analyses, the latter illustrated by \cite{sensitivity} where it is performed for a model which discriminates reactor types based on measurements of plutonium samples and their errors.

This problem has been addressed in the past through pre-computed data-bases of reactor simulations, and the interpolation of either the one-group cross-sections \cite{arp} \cite{cyclus2} or nuclide concentrations resulting from these simulations with different methods such as: Nearest Neighbors, Linear Fit, Lagrange Polynomials, Neural Networks \cite{baptiste}, and Cubic Splines, the latter being currently the most commonly used.

Up until now, most of the available software packages have used infinite lattice reactor simulations sampled on a uniform multi-dimensional grid of at most three dimensions \cite{arp}\cite{cyclus2}. For low dimensional problems ($d\leq3$), uniform grid sampling produces good results, however for a larger number of dimensions, this sampling method does not distribute the sampling points efficiently across space \cite{DOE}, resulting in poor exploration of the parameter space. Quasirandom Sampling methods would provide better space coverage properties at higher dimensions \cite{DOE}. However, some of the above mentioned interpolation methods require the samples to be distributed on a grid in order to perform correctly. Furthermore, while the interpolation quality can be estimated, the aforementioned interpolation methods do not provide information on the expected variance of the interpolation at non-sampled points. 

Here, we propose using Gaussian Processes (GP) for interpolation, by considering noiseless inputs. Under this assumption, GP's provide zero error on points arising from the dataset on which the models have been trained, whereas a regression is performed for unseen data points. In addition, this method has already been used in conjunction with Quasirandom Sampling \cite{ESARDA-Antonio}. Furthermore, also the gradient and estimates of their prediction uncertainty at non-sampled points are directly obtained. The former is useful for applications such as forward and inverse optimization problems, the latter for uncertainty quantification applications. GP belongs to a set of tools used in the Machine Learning communities for a variety of tasks, including classification and regression. With Gaussian Process Regression (GPR), the interpolation is not performed on a specific function but over an infinite distribution of functions that share common properties as defined by the user.

An equivalent concept to GPR called kriging is well known in the field of geology \cite{recipes}. Recently, researchers have used GP-based surrogate models for nuclear engineering applications such as modeling of equipment degradation and preventive maintenance \cite{degradation}, or studying fuel performance and thermo-hydraulics \cite{BayesGP}. They also have been used for the prediction of fuel nuclide compositions and compared to surrogate models based on Dynamic
Mode Decomposition, by performing regression on a Principal Component Analysis (PCA) model of the fuel isotopics \cite{dmd}. Additionally, we have started exploring  their use for the direct prediction of spent fuel compositions, without a PCA-reduced model and using multidimensional input variables \cite{ESARDA-Antonio}, work that we update here

In this paper, we compare the performance of GP-based surrogate models to models based on Cubic Splines for the direct interpolation of spent fuel nuclide compositions (GP could also interpolate cross-sections). We compare to Cubic Splines as they often produce smoother and higher quality interpolators in comparison to methods such as Langrange and Newton polynomials \cite{na}. The performance of both techniques will be explored through computational experiments involving different sampling configurations using both Grid sampling and Sobol Quasirandom Sequences in order to assess the impact of the sampling strategy on the regressions. We study two-dimensional problems to examine whether GPR performs well already in problems where grid sampling is still effective.

%% file: Sections/TheoreticalBackground.tex
\section{Creating the datasets}\label{sec:Models}
\subsection{Reactor simulations}
\begin{figure}
    \centering
    \includegraphics[width=0.5\textwidth]{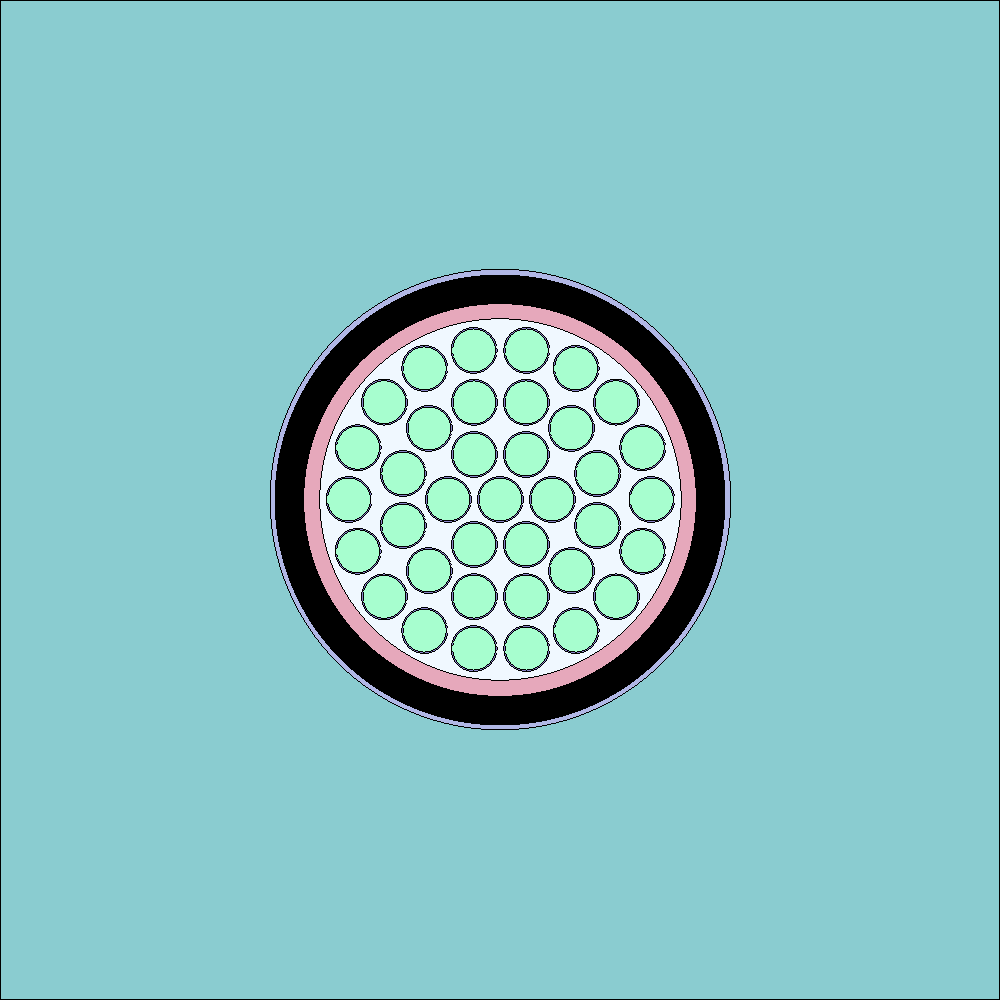}
    \caption{CANDU 6, 37-element fuel assembly, 2D infinite lattice implementation in \texttt{SERPENT 2} (light green indicates the fuel elements, white for the coolant, pink for the calandria and pressure tubes, black for the void between these tubes and turquoise for the moderator between fuel channels)}
    \label{fig:2dlattice}
\end{figure}

For our research, we have modeled a 2D infinite lattice model of a CANDU 6 reactor based on specifications from available literature \cite{Candu}. The implementation has been made in the computer code \texttt{SERPENT 2} which couples a Monte Carlo neutron transport module with a fuel depletion solver based on the Chebyshev Rational Approximation Method \cite{serpent}. The quality of the model has been examined by comparing end-of-cycle isotopic compositions with the \texttt{Bruce-1} dataset reported in the SFCOMPO-2.0 database \cite{sfcompo}. Figure \ref{fig:2dlattice} shows the CANDU 6 37-elements fuel assembly implemented in \texttt{SERPENT 2}.

\begin{table}[ht]
    \centering
    \begin{tabular}{c c}
        \toprule
        \textbf{Parameters} & \textbf{Range} \\ \midrule
         Moderator Temperature & 333 - 363 $K$ \\
         Burnup & 0.1 - 7 $\frac{\mathrm{MWd}}{\mathrm{kg_{HM}}}$ \\
         \bottomrule
    \end{tabular}
    \caption{Parameter ranges used in in the generation of the training and testing datasets}
    \label{tab:paramranges}
\end{table}

\subsection{Sampling strategies}
Strategies are required to sample sets of input parameters for each Serpent simulation involved in model development and testing. In this study, we sample moderator temperature and discharge burnup. Table \ref{tab:paramranges} indicates the range of values considered for these parameters. 

Previous studies on sampling schemes and strategies have shown that uniform grid sampling performs poorly in higher-dimensional input spaces \cite{DOE}. While a random number sequence can overcome this issue in principle, due to the nature of pseudo-random number generators implementations, random number sequences tend to cluster, resulting in a non-uniform distribution of samples across the input space. In such cases, a quasirandom sequence can provide a set of samples with a better spatial distribution\cite{qsr}

Sobol quasirandom sampling is a method to generate quasi-random sequences which is designed to minimize the star discrepancy, a mathematical term describing the difference between the distribution of values generated by a certain generating scheme to a multidimensional uniform probability distribution. The generation of the samples involves using a special algorithm in which bitwise operations are performed. Details on the algorithm and its implementation can be found in \cite{SobolSQ}.

\begin{figure}
    \centering
    \input{Sections/Images/sampling_comparison}
    \caption{Comparison between the dataset based on Sobol quasirandom sampling (left) and the dataset based on Grid sampling (right). While both the Sobol sequence and grid sampling are effective in two dimensions as seen here, the Sobol sequence by far outperforms grid sampling in higher dimensions, at the same number of samples.}
    \label{fig:SvG}
    
\end{figure}
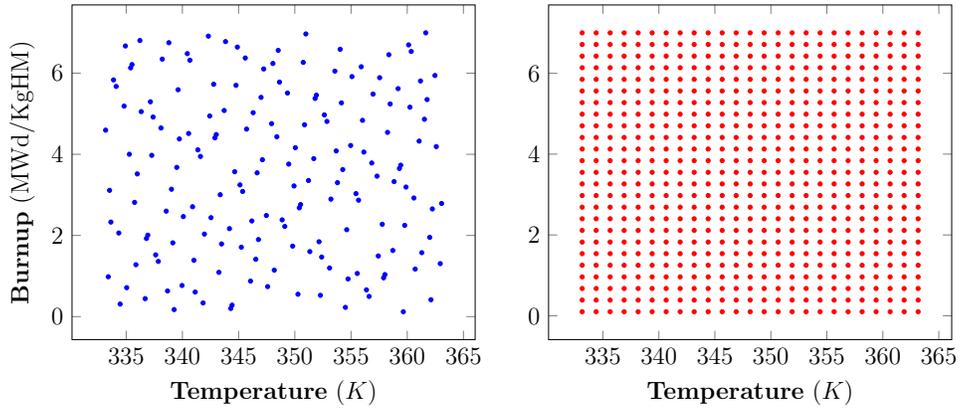

\begin{figure}
    \centering
    \includegraphics[width=0.32\textwidth,trim=2cm 1.3cm 2cm 1.4cm, clip]{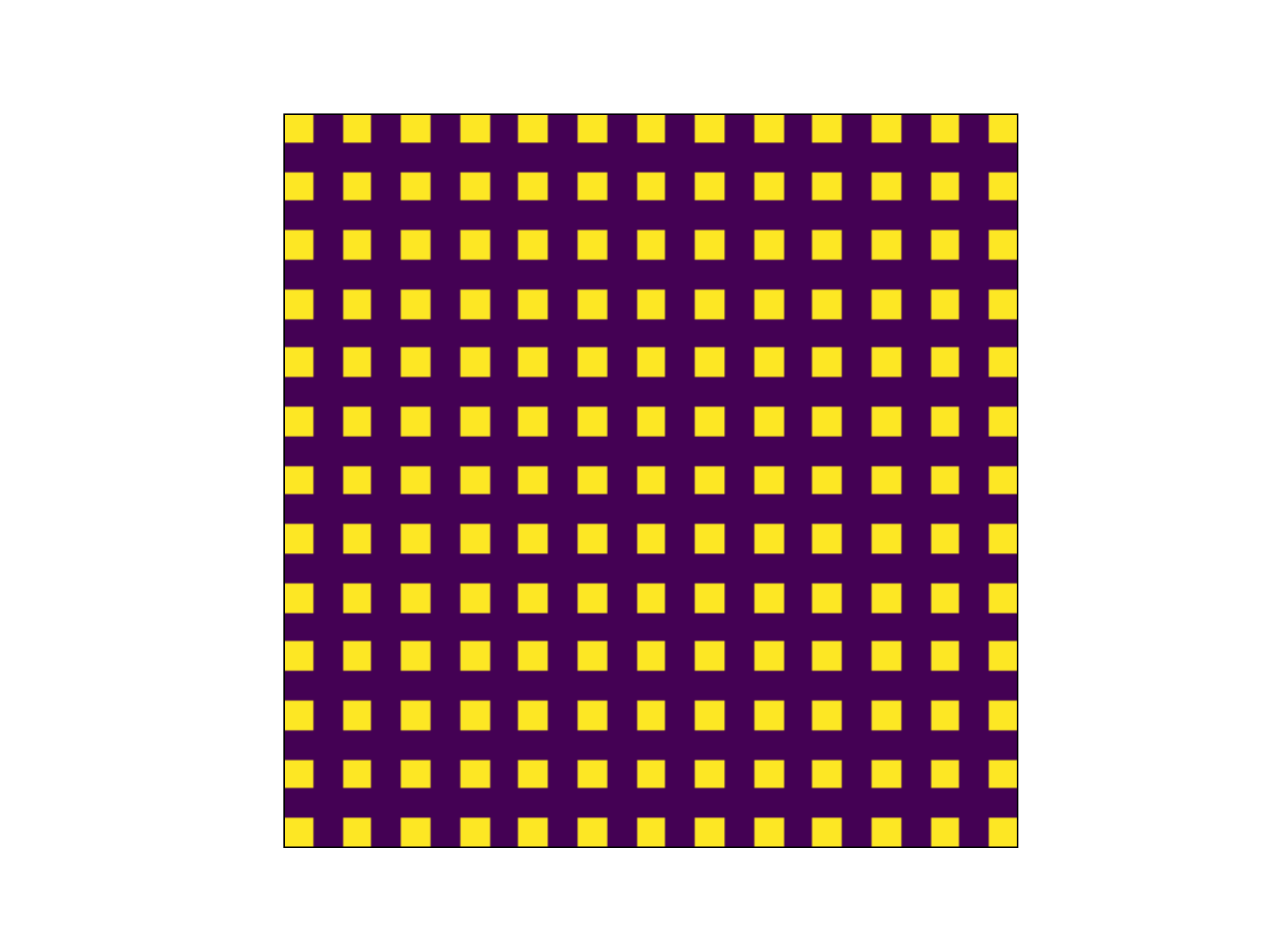}
    \caption{Spatial distribution pattern used to evaluate the performance of the surrogate models on a grid. The yellow elements represent the simulations to be used for training the interpolators, while the dark regions represent the cells where the values will be interpolated.
    }
    \label{fig:exppatterns}
\end{figure}

In order to evaluate the impact of the sampling method on the prediction quality of the models, two scenarios have been prepared.  In Scenario A, a 25x25 2D-grid dataset was used and simulated, based on the parameter ranges of Table \ref{tab:paramranges}. In this scenario a set of 169 samples was generated from the selection of samples starting from the corners and leaving one grid element in between. The rest of the unselected grid points are then considered as the test dataset, for both this and the following scenario. Scenario B consists of 169 samples generated from a Sobol sequence sampler written in-house. Again, the test dataset is the same as for Scenario A. \\ Figure \ref{fig:SvG} shows a comparison between the samples generated via the Sobol sequence and the aforementioned grid. While grid sampling is still effective in two dimensions, the good space coverage of the Sobol sequence is evident. Figure \ref{fig:exppatterns} shows the spatial arrangement of Scenario A's training set through the bright elements, while the darker elements represent the test dataset used for both scenarios. Table \ref{tab:setups} summarizes the different scenarios and their characteristics

\begin{table}[ht]
    \centering
    \begin{tabular}{c c c}
    \toprule
     \textbf{Configuration}&\textbf{Training set size}  & \textbf{Test set size}   \\ \midrule
     Scenario A& 169 (Grid)  & 456 (Grid)   \\
     Scenario B & 169 (Sobol) & 456 (Grid) \\
     
    \bottomrule 
    \end{tabular}
    \caption{Details of the sampling strategies. The different configurations allow for the study of model performance under varying spatial sample distributions}
    \label{tab:setups}
\end{table}

\section{Building the model}
\label{sec:Reg}

Based on the pre-generated training data, we built the Cubic Spline and the GPR models. In the following section we examine these methods.

\subsection{Cubic Spline Interpolation}
Cubic Spline Interpolation is a method to interpolate functions within the boundaries described by a number of its samples. The interpolation procedure consists of fitting a piecewise cubic polynomial between the sampled points. This method results in smoother and higher quality interpolators in comparison to methods such as Langrange and Newton polynomial fitting \cite{NM2}. In addition, it has the advantage of avoiding oscillations and Runge's phenomenon. A more detailed description of the method and its implementation can be found in \cite{CSplines}.


\subsection{Gaussian Process Regression}

A GP is a set of random variables that share a joint Gaussian probability distribution. GP's can be used for the construction of probabilistic surrogate models of black-box problems where an analytical form is unavailable or intractable and the computational cost is elevated. Without loss of generalization, the aim of GPR is to approximate a function $Y(\vv{X}),\, \vv{X} = (x_{1},..., x_{d})$ through a GP as:
\begin{equation}
    Y(\vv{X}) \approx \mathcal{G}\mathcal{P}(\vv{X}) = \mathcal{N}(0,K(\vv{X}))
\end{equation}
Where $d$ is the dimension of the input space. The kernel ($K$) of a GP is a function which describes the covariance between the inputs of the target function. It encodes a limited set of assumptions about the underlying function such as smoothness and differentiability. By choosing a kernel, an infinite set of functions which share the kernel properties are used to perform the regression, and by training the kernel parameters based on the input and output data a subset of those functions are chosen that match the data.
We chose the Anisotropic Squared Exponential (ASE) kernel, which provides for very smooth interpolation and is infinitely differentiable, as we expect that the change of nuclide concentrations throughout the parameter space would meet these characteristics. An additional advantage is that the ASE kernel allows for the determination of relative input parameter relevance through the use of different correlation lengths parameters ($\ell_{i}$) for each input:

\begin{equation}\label{eq:ASE}
    K(\vv{X},\vv{Z}) = exp\left(-\sum_{i}^{d}\left( \dfrac{x_{i}-z_{i} } {\ell_{i}}\right)^{2}\right)
\end{equation}
Where $\vv{X},\vv{Z}$ are two different points in which the covariance function is evaluated. The smaller $\ell_{i}$, the more sensitive the underlying model is to changes in input $x_{i}$. Once the parameters of the kernel that reproduce the training data are estimated, the posterior predictive distribution of the GP - our model - at an unseen point ($*$) is given by a normal distribution with prediction mean ($\mu_{*}$), prediction variance ($\sigma^{2}_{*}$) and prediction gradient ($\nabla\mu_{*}$):
\begin{align}
    \mu_{*} &= K_{train,*}  (K_{train,train})^{-1} Y_{train} \nonumber \\
    \sigma^{2}_{*} &= K_{*,*} - K_{*,train}^{T}(K_{train,train})^{-1}K_{*,train} \nonumber \\
    \nabla\mu_{*} &= \nabla K_{train,*}  (K_{train,train})^{-1} Y_{train}
\end{align}
    
 To implement the GPR models, we have used the \texttt{scikit-learn} Python package \cite{scikit-learn}. More details on GP's and their kernels can be found in \cite{6}. 


\subsection{Cross-validation}

Since the kernel training process is strongly dependent on the training data, weak prediction performance can occur if an inappropriate selection of the training set is made. This can be avoided by cross-validation.
It is implemented by splitting the training set into k ``folds'' of approximately equal size and performing the training on each fold, thus obtaining the model parameters, then using the other folds as test data to evaluate the model predictive quality. 

This should be performed several times by randomizing the selection of the folds, thus generating a set of plausible model parameters from which the best performing combinations can be selected. Notwithstanding this, a major benefit of cross-validation is that it allows the training of models with different combinations of samples spanning the entire input space, providing parameter sets that tend to enhance the model generalization properties, typically at a smaller computational cost since the training set size is reduced. 

\begin{figure}
    \centering
    \resizebox{.95\textwidth}{!}{
    \input{Sections/Images/Pu2d}
    \input{Sections/Images/Pu1d}
    }
    \caption{Reconstruction plots for $^{239}$Pu. The left plot shows the GPR model based on samples generated by the Sobol sequence, showing the mass as a function of burnup and temperature. The reported masses have been scaled by the number of fuel assemblies in the reactor and their dimensions. The squares indicate the samples obtained from the Sobol sequence. The plot on the right shows the same model but plotting this time the predictions for two fixed temperature values alongside their posterior uncertainties. The latter are shown at the 6-$\sigma$ level. 
    	A portion of the high burnup section of the plot has been zoomed in in the top left corner, making evident the small albeit important effect of temperature at higher burnups. A lower temperature results in more plutonium being produced. This can be explained by a thermal broadening of $^{238}$U resonances. Below this plot, the errors of both curves is shown. It can be observed that they are distributed around zero.}
    \label{fig:gpr_model}
\end{figure}
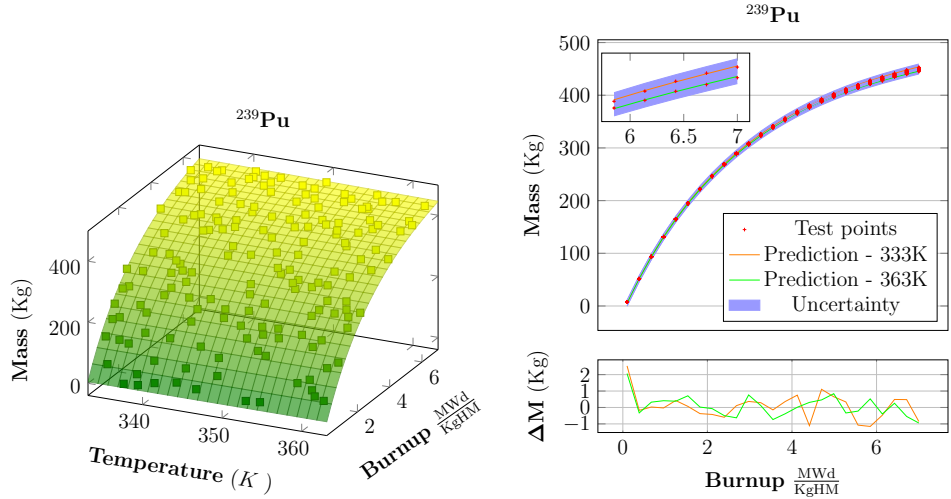

Each \texttt{SERPENT 2} simulation provides an output vector containing about 1300 nuclides. We have created GPR models for each of these nuclides for both datasets. Each GPR model has been trained using a 5-fold cross validation scheme that has been repeated 10 times, thus generating a set of 50 kernel parameter combinations from which the best-performing is chosen. Figures \ref{fig:gpr_model}, \ref{fig:sr90} and \ref{fig:eu154} show the computed GPR models for $^{239}$Pu, $^{90}$Sr and $^{154}$Eu respectively. The models shown are based on the Sobol sequence dataset from scenario B.

%% file: Sections/Images/sampling_comparison.tex
\pgfplotsset{select coords between index/.style 2 args={
		x filter/.code={
			\ifnum\coordindex<#1\def\pgfmathresult{}\fi
			\ifnum\coordindex>#2\def\pgfmathresult{}\fi
		}
}}
\resizebox{.95\textwidth}{!}{
\begin{tikzpicture}
\begin{axis}[name = sobol,xlabel = \textbf{Temperature} ($K$),ylabel=\textbf{Burnup} (MWd/KgHM)]
\addplot[blue, only marks,mark options={scale=0.5} ,select coords between index={0}{169}]
table [x=/*temperature*/, y=/*burnup*/, col sep=comma] {Sections/Images/A.csv};
\end{axis}
\begin{axis}[name = grid, xshift = 1cm, at = (sobol.right of south east), xlabel = \textbf{Temperature} ($K$)]
\addplot[red, only marks,mark options={scale=0.5}]
table [x=/*temperature*/, y=/*burnup*/, col sep=comma] {Sections/Images/Grid.csv};
\end{axis}
\end{tikzpicture}
}

%% file: Sections/Images/Pu2d.tex
	\newcommand{\dsurfPu}{Sections/Images/2dSobolPu239.csv}
\newcommand{\dsurfPuN}{Sections/Images/Pu2392dComplement.csv}
    \begin{tikzpicture}	
\begin{axis}[name = surf,
	smooth,
	legend pos = north east,
	colormap/greenyellow,
	xlabel = {\textbf{Temperature} ($K$ )},
	ylabel = \textbf{Burnup} $\frac{\mathrm{MWd}}{\mathrm{KgHM}}$,
	zlabel = \textbf{Mass} (Kg),
	xlabel style={rotate=-9},
	ylabel style={rotate=35},  title= $^{239}$\textbf{Pu}
	]
	\addplot3 [surf,z buffer=sort, mesh/rows=25, opacity=0.6]
	table[y={Burnup}, x={Temperature} , z={Pred}, col sep=comma] {\dsurfPu};
	\addplot3+[only marks, scatter] table[y={Burnup}, x={Temperature} , z={Yts}, col sep=comma] {\dsurfPuN};
\end{axis} %
	\end{tikzpicture}

%% file: Sections/Images/Pu1d.tex
\newcommand{\dataPu}{Sections/Images/1dSobolPu239GPR.csv}
	\begin{tikzpicture}[scale=1]
	\pgfkeys{/tikz/savenumber/.code 2 args={\global\edef#1{#2}}}
    	\begin{axis}[name = collapsed,
	ylabel=\textbf{Mass} (Kg),
	grid=both,
	xtick = \empty,
	legend pos=south east,title= $^{239}$\textbf{Pu}]
	\addplot[red,only marks,mark size=1pt, mark = +] table[x={Burnup},y={Yts},col sep=comma]{\dataPu};
	\addplot[orange,thin,select coords between index={0}{24}] table [x={Burnup}, y={Pred},col sep=comma]{\dataPu};
	\addplot[green,thin,select coords between index={600}{624}] table [x={Burnup}, y={Pred},col sep=comma]{\dataPu};
	
		\addplot[name path=A,blue!40,thin,select coords between index={0}{24}] table [x={Burnup}, 	y expr=\thisrow{Pred}+6*\thisrow{Std},col sep=comma]{\dataPu};
	\addplot[name path=B,blue!40,thin,select coords between index={0}{24}] table [x={Burnup}, 	y expr=\thisrow{Pred}-6*\thisrow{Std},col sep=comma]{\dataPu};
	\addplot[blue!40] fill between[of=A and B];
			\addplot[name path=C,blue!40,thin,select coords between index={600}{624}] table [x={Burnup}, 	y expr=\thisrow{Pred}+6*\thisrow{Std},col sep=comma]{\dataPu};
	\addplot[name path=D,blue!40,thin,select coords between index={600}{624}] table [x={Burnup}, 	y expr=\thisrow{Pred}-6*\thisrow{Std},col sep=comma]{\dataPu};
	\addplot[blue!40] fill between[of=C and D];
	\legend{Test points,Prediction - 333K,Prediction - 363K,,,Uncertainty};
	
	   \coordinate (insetSW) at (axis cs:-0.5,350); 
    \coordinate (insetNE) at (axis cs:3,480); 
      \path
      let
        \p1 = (insetSW),
        \p2 = (insetNE),
        \n1 = {(\x2 - \x1)},
        \n2 = {(\y2 - \y1)}
      in    
      [savenumber={\insetwidth}{\n1},savenumber={\insetheight}{\n2}];
	\end{axis}
		\begin{axis}[at={(insetSW)},anchor = south west, name=inset, width=\insetwidth,
    height=\insetheight, scale only axis, ytick=\empty] 

	\addplot[red,only marks, mark size=1pt, mark = +,select coords between index={20}{24}] table [x={Burnup}, y={Yts},col sep=comma]{\dataPu};
	\addplot[red,only marks, mark size=1pt, mark = +,select coords between index={620}{624}] table [x={Burnup}, y={Yts},col sep=comma]{\dataPu};
	\addplot[orange,thin,select coords between index={20}{24}] table [x={Burnup}, y={Pred},col sep=comma]{\dataPu};
	\addplot[green,thin,select coords between index={620}{624}] table [x={Burnup}, y={Pred},col sep=comma]{\dataPu};
			\addplot[name path=A,blue!40,thin,select coords between index={20}{24}] table [x={Burnup}, 	y expr=\thisrow{Pred}+6*\thisrow{Std},col sep=comma]{\dataPu};
	\addplot[name path=B,blue!40,thin,select coords between index={20}{24}] table [x={Burnup}, 	y expr=\thisrow{Pred}-6*\thisrow{Std},col sep=comma]{\dataPu};
	\addplot[blue!40] fill between[of=A and B];
			\addplot[name path=C,blue!40,thin,select coords between index={620}{624}] table [x={Burnup}, 	y expr=\thisrow{Pred}+6*\thisrow{Std},col sep=comma]{\dataPu};
	\addplot[name path=D,blue!40,thin,select coords between index={620}{624}] table [x={Burnup}, 	y expr=\thisrow{Pred}-6*\thisrow{Std},col sep=comma]{\dataPu};
	\addplot[blue!40] fill between[of=C and D];
	\end{axis}
	\begin{axis}[name = errors,
	at = (collapsed.south west),
	yshift = -2cm,
	ylabel = $\bf{\Delta}$\textbf{M} (Kg),
	yscale=0.25,
	grid = both,
	xlabel =\textbf{Burnup} $\frac{\mathrm{MWd}}{\mathrm{KgHM}}$, every axis x label/.style={at={(0.5,-3)}}]
    \addplot[orange, select coords between index={0}{24}] table [x={Burnup}, 	y expr=\thisrow{Yts}-\thisrow{Pred},col sep=comma]{\dataPu};
	\addplot[green, select coords between index={600}{624}] table [x={Burnup}, 	y expr=\thisrow{Yts}-\thisrow{Pred},col sep=comma]{\dataPu};
    \end{axis}
	\end{tikzpicture}

%% file: Sections/Results.tex
\section{Results and Discussion}\label{section:results}
We have compared GPR-based models of spent fuel nuclide concentrations to Cubic Splines models of these quantities. The Splines models have been implemented using the \texttt{SmoothBivariateSpline} module from the \texttt{Scipy} \cite{scipy} python package. The following comparison is focused on two main elements: Runtime Performance and Model Quality. 

\subsection{Runtime performance}
We have studied the mean time required to perform a prediction for the interpolation methods discussed in this article. This calculation time we refer to here as \textit{runtime} is different from the time required to train the model which in itself is dependent on the sample size, the number of dimensions and the subsequent optimization problem that is to be solved in order to derive the appropriate kernel hyperparameters. The tests have been performed on a 2.2 GHz \texttt{Intel Core i7} computer with 4 cores and 16GB of RAM. Table \ref{tab:runtime} shows the regression time required averaged over all the models, with the time for a GP model separated into the time required for the prediction \textbf{GPR}$-\mu$ and the time required for the estimation of the prediction variance \textbf{GPR}$-\sigma$, as these calculations can be run independently of each other.

Interpolation using GP models can require significantly more time than interpolation based on Cubic Splines. It might be possible that through the use of more efficient implementations, the runtime difference between the two can be reduced in the future.

\input{Sections/Tables/tableruntime}

\subsection{Model Quality}
We have analyzed the quality of the models based on GP's and Cubic Splines using the experimental designs presented in Table \ref{tab:setups} and Figures \ref{fig:SvG} and \ref{fig:exppatterns}. For this, we have considered an array of metrics to quantify model performance, namely the root-mean-square-errors ($RMSE$) and the coefficients of determination $R^{2}$. We also examine an additional metric related to the posterior predictive variance produced by GPR in order to study the quality of this estimator. This quantity is $Pred-1\sigma$, which quantifies the fraction of model predictions located within 1 predictive standard deviation from the true values of the test sets, i.e. it shows whether the predicted variance represents the variance that is actually observed. 

While models were created for each of the nuclides tracked by \texttt{SERPENT 2}, Tables 
\ref{tab:grid2} and \ref{tab:sobolnew} contain the results of
our analysis for a small subset of them representative of both major actinides and fission products. The mean values shown have been scaled by the assembly length and number of fuel assemblies in the reactor core. 
\input{Sections/Tables/tablegrid2}
\input{Sections/Tables/tablesobolnew}

In general, we observe that based on the $R^{2}$ metric, both interpolation methods have good performance for most of the nuclides. Furthermore, we note that based on the $RMSE$ metric, GP based models perform  better in all the experimental configurations for all isotopes. The improvement of the mean RMSE is in most cases between a factor of 1.02 and 10. Still, Cubic Splines could provide on average reasonable results for most nuclides, provided the application requirements allow for $RMSE$ errors of such magnitude.

We also observe that the variance predictor of the GP model predicts the true values at the $1-\sigma$ level reasonably well, which is not generalizable to any GPR application. In addition,it can be seen that both surrogate model methods perform better on a uniform grid-based sample set compared to a sampled set based on Sobol sequences. Nevertheless, based on a study where different sampling schemes are compared for their space covering properties and efficiency \cite{DOE}, one would expect the positive properties of Sobol sequences to be made manifest in problems of higher dimensionality.


         
     

%% file: Sections/Tables/tableruntime.tex
\begin{table*}[ht]
    \centering
    \begin{tabular}{c c c c c}\toprule
    \textbf{Configuration} & \textbf{Cubic Spline}  & \textbf{GPR}$-\mu$ & \textbf{GPR}$-\sigma$ & \textbf{GPR Total} \\
    & [s] &[s] &[s] &[s] \\ \midrule
   Scenario A & 3.40e-05 & 4.61e-04 & 6.00e-04 & 1.06e-03  \\
   Scenario B & 9.74e-06 & 1.18e-03& 9.95e-04& 2.18e-03 \\
   \bottomrule
     
    \end{tabular}
    \caption{Run-time comparison for different configurations.} 
    \label{tab:runtime}
\end{table*}

%% file: Sections/Tables/tablegrid2.tex
\begin{center}
\begin{table*}[ht!]
\centering
	\resizebox{\tablesize\textwidth}{!}{
		\begin{filecontents*}{Sections/Tables/Even_all_2space.csv}
Nuclide,mean_isotope,Spline-RMSE,GPR-RMSE,RatioCubicGPR,SplineRsquared,GPRRsquared,GPRPredictionin1sigma,GPRPredictionin2sigma
$^{85}\mathrm{Rb} $,2.0561483224704005,$3.9\cdot 10^{-3}$,\cellcolor{Gray}$3.4\cdot 10^{-3}$,1.1399617589613085,$1.2\cdot 10^{-5}$,$9.8\cdot 10^{-6}$,73.68421052631578,91.00877192982456
$^{90}\mathrm{Sr} $,11.5504532608896,$2.3\cdot 10^{-2}$,\cellcolor{Gray}$1.9\cdot 10^{-2}$,1.1993298489575115,$1.4\cdot 10^{-5}$,$1.0\cdot 10^{-5}$,73.90350877192982,91.66666666666666
$^{106}\mathrm{Ru} $,4.3397494121088,$1.8\cdot 10^{-2}$,\cellcolor{Gray}$7.8\cdot 10^{-3}$,2.365458394268485,$3.8\cdot 10^{-5}$,$6.8\cdot 10^{-6}$,75.21929824561403,90.13157894736842
$^{110}\mathrm{Cd} $,0.11709530187192958,$4.8\cdot 10^{-4}$,\cellcolor{Gray}$3.2\cdot 10^{-4}$,1.4909254366487843,$1.6\cdot 10^{-5}$,$7.2\cdot 10^{-6}$,68.85964912280701,84.21052631578947
$^{135}\mathrm{Cs} $,2.5784273234688,$5.0\cdot 10^{-3}$,\cellcolor{Gray}$4.4\cdot 10^{-3}$,1.1513743635879314,$1.1\cdot 10^{-5}$,$8.3\cdot 10^{-6}$,75.0,91.44736842105263
$^{137}\mathrm{Cs} $,24.476815362432,$4.2\cdot 10^{-2}$,\cellcolor{Gray}$4.1\cdot 10^{-2}$,1.0164393031940921,$8.8\cdot 10^{-6}$,$8.5\cdot 10^{-6}$,65.35087719298247,83.99122807017544
$^{137}\mathrm{Ba} $,0.16981773525987842,$3.4\cdot 10^{-4}$,\cellcolor{Gray}$3.3\cdot 10^{-4}$,1.0151482093824358,$5.2\cdot 10^{-6}$,$5.0\cdot 10^{-6}$,98.46491228070175,99.56140350877193
$^{143}\mathrm{Nd} $,16.524794392354558,$2.2\cdot 10^{-1}$,\cellcolor{Gray}$3.0\cdot 10^{-2}$,7.3236233149625765,$5.2\cdot 10^{-4}$,$9.7\cdot 10^{-6}$,66.00877192982456,90.5701754385965
$^{147}\mathrm{Nd} $,0.847539624768,$6.8\cdot 10^{-2}$,\cellcolor{Gray}$8.8\cdot 10^{-3}$,7.768293861322435,$3.3\cdot 10^{-1}$,$5.5\cdot 10^{-3}$,64.69298245614034,95.83333333333334
$^{148}\mathrm{Nd} $,7.563249000537599,$1.3\cdot 10^{-2}$,\cellcolor{Gray}$1.2\cdot 10^{-2}$,1.019502281979532,$9.0\cdot 10^{-6}$,$8.7\cdot 10^{-6}$,75.87719298245614,92.54385964912281
$^{147}\mathrm{Sm} $,0.4961471924369127,$1.1\cdot 10^{-3}$,\cellcolor{Gray}$1.0\cdot 10^{-3}$,1.11052779329682,$6.4\cdot 10^{-6}$,$5.2\cdot 10^{-6}$,77.19298245614034,92.32456140350878
$^{151}\mathrm{Sm} $,0.29347478954879996,$1.3\cdot 10^{-2}$,\cellcolor{Gray}$6.8\cdot 10^{-4}$,19.89245859425107,$2.7\cdot 10^{-2}$,$7.0\cdot 10^{-5}$,100.0,100.0
$^{154}\mathrm{Eu} $,0.12755407819212286,$2.8\cdot 10^{-4}$,\cellcolor{Gray}$2.6\cdot 10^{-4}$,1.0825520394293204,$6.7\cdot 10^{-6}$,$5.7\cdot 10^{-6}$,82.01754385964912,96.05263157894737
$^{239}\mathrm{Pu} $,301.75855186944,$1.0\cdot 10^{0}$,\cellcolor{Gray}$5.2\cdot 10^{-1}$,1.9547699856987646,$6.5\cdot 10^{-5}$,$1.7\cdot 10^{-5}$,63.1578947368421,92.54385964912281
$^{240}\mathrm{Pu} $,75.1537166934144,$5.8\cdot 10^{-1}$,\cellcolor{Gray}$1.4\cdot 10^{-1}$,4.083304294792766,$1.1\cdot 10^{-4}$,$6.6\cdot 10^{-6}$,74.3421052631579,90.5701754385965
$^{241}\mathrm{Pu} $,13.602905556247832,$5.4\cdot 10^{-2}$,\cellcolor{Gray}$3.3\cdot 10^{-2}$,1.6435218568066507,$2.1\cdot 10^{-5}$,$8.0\cdot 10^{-6}$,74.12280701754386,89.47368421052632
$^{241}\mathrm{Am} $,0.11377797377158645,$1.8\cdot 10^{-3}$,\cellcolor{Gray}$3.1\cdot 10^{-4}$,6.007601777188097,$2.3\cdot 10^{-4}$,$6.5\cdot 10^{-6}$,77.63157894736842,91.00877192982456
\end{filecontents*}
\pgfplotstabletypeset[
		multicolumn names, 
		col sep=comma,
		columns = {Nuclide,mean_isotope,Spline-RMSE,GPR-RMSE,RatioCubicGPR,SplineRsquared,GPRRsquared,GPRPredictionin1sigma},
		columns/Nuclide/.style={
			column name=$Nuclide$, 
			string type},  
		columns/mean_isotope/.style={
			column name=$\mu$, 
			precision=2,fixed}, 
		columns/Spline-RMSE/.style={
			column name= Cubic Spline,
			string type},
		columns/GPR-RMSE/.style={
			column name=$GPR$,string type},
		columns/RatioCubicGPR/.style={
		column name=RMSE $\frac{Spline}{GPR}$,
		precision=2},
		columns/SplineRsquared/.style={
			column name=$1 - R^{2}_{Sp}$,
			string type},
		columns/GPRRsquared/.style={
			column name=$1 - R^{2}_{GPR}$,
			string type},
		columns/GPRPredictionin1sigma/.style={
			column name=$Pred-1\sigma$,
			precision=1,},
		every head row/.style={
			before row={\toprule}, 
			after row={
							&	Kg& $\mathrm{RMSE}$ & $\mathrm{RMSE}$  & &  & & $\%$ \\
				\midrule} 
		},
		every last row/.style={after row=\bottomrule}, 
		]{Sections/Tables/Even_all_2space.csv} 
		}
			\caption{\textbf{Scenario A:} Comparison between GPR and Cubic spline models for a selection of nuclides. Highlighted cells indicate smallest error in the comparison. RMSE values are in Kg}
			\label{tab:grid2}
\end{table*}
\end{center}

%% file: Sections/Tables/tablesobolnew.tex
\begin{center}
\begin{table*}[ht!]
\centering
	\resizebox{\tablesize\textwidth}{!}{
		\begin{filecontents*}{Sections/Tables/SobolNew169_Sobol_all_2space.csv}
Nuclide,mean_isotope,Spline-RMSE,GPR-RMSE,RatioCubicGPR,SplineRsquared,GPRRsquared,GPRPredictionin1sigma,GPRPredictionin2sigma
$^{85}\mathrm{Rb} $,2.0561483224704005,$1.0\cdot 10^{-2}$,\cellcolor{Gray}$6.1\cdot 10^{-3}$,1.6493220534489625,$8.3\cdot 10^{-5}$,$3.0\cdot 10^{-5}$,100.0,100.0
$^{90}\mathrm{Sr} $,11.5504532608896,$5.8\cdot 10^{-2}$,\cellcolor{Gray}$3.3\cdot 10^{-2}$,1.7407180389730577,$9.3\cdot 10^{-5}$,$3.0\cdot 10^{-5}$,96.0,100.0
$^{106}\mathrm{Ru} $,4.3397494121088,$2.8\cdot 10^{-2}$,\cellcolor{Gray}$2.3\cdot 10^{-2}$,1.2177932630270512,$8.6\cdot 10^{-5}$,$5.8\cdot 10^{-5}$,100.0,100.0
$^{110}\mathrm{Cd} $,0.11709530187192958,$1.4\cdot 10^{-3}$,\cellcolor{Gray}$7.4\cdot 10^{-4}$,2.008390765336926,$1.4\cdot 10^{-4}$,$3.5\cdot 10^{-5}$,100.0,100.0
$^{135}\mathrm{Cs} $,2.5784273234688,$1.1\cdot 10^{-2}$,\cellcolor{Gray}$7.7\cdot 10^{-3}$,1.50886027254812,$5.6\cdot 10^{-5}$,$2.4\cdot 10^{-5}$,100.0,100.0
$^{137}\mathrm{Cs} $,24.476815362432,$1.1\cdot 10^{-1}$,\cellcolor{Gray}$4.0\cdot 10^{-2}$,2.744723150855203,$5.9\cdot 10^{-5}$,$7.8\cdot 10^{-6}$,92.96,99.83999999999999
$^{137}\mathrm{Ba} $,0.16981773525987842,$1.4\cdot 10^{-3}$,\cellcolor{Gray}$3.4\cdot 10^{-4}$,4.242593046528721,$9.3\cdot 10^{-5}$,$5.1\cdot 10^{-6}$,100.0,100.0
$^{143}\mathrm{Nd} $,16.524794392354558,$2.0\cdot 10^{-1}$,\cellcolor{Gray}$5.8\cdot 10^{-2}$,3.444983468750733,$4.2\cdot 10^{-4}$,$3.5\cdot 10^{-5}$,90.56,93.92
$^{147}\mathrm{Nd} $,0.847539624768,$7.6\cdot 10^{-2}$,\cellcolor{Gray}$3.9\cdot 10^{-3}$,19.65668180975611,$3.1\cdot 10^{-1}$,$8.1\cdot 10^{-4}$,100.0,100.0
$^{148}\mathrm{Nd} $,7.563249000537599,$3.3\cdot 10^{-2}$,\cellcolor{Gray}$1.4\cdot 10^{-2}$,2.3076461240187434,$5.6\cdot 10^{-5}$,$1.0\cdot 10^{-5}$,100.0,100.0
$^{147}\mathrm{Sm} $,0.4961471924369127,$3.9\cdot 10^{-3}$,\cellcolor{Gray}$1.8\cdot 10^{-3}$,2.1596441429042095,$7.8\cdot 10^{-5}$,$1.6\cdot 10^{-5}$,100.0,100.0
$^{151}\mathrm{Sm} $,0.29347478954879996,$1.5\cdot 10^{-2}$,\cellcolor{Gray}$1.5\cdot 10^{-3}$,10.114269549983323,$3.4\cdot 10^{-2}$,$3.3\cdot 10^{-4}$,100.0,100.0
$^{154}\mathrm{Eu} $,0.12755407819212286,$9.9\cdot 10^{-4}$,\cellcolor{Gray}$4.0\cdot 10^{-4}$,2.4431223581611174,$7.8\cdot 10^{-5}$,$1.3\cdot 10^{-5}$,100.0,100.0
$^{239}\mathrm{Pu} $,301.75855186944,$2.2\cdot 10^{0}$,\cellcolor{Gray}$9.8\cdot 10^{-1}$,2.2789208892613075,$3.0\cdot 10^{-4}$,$5.9\cdot 10^{-5}$,9.120000000000001,17.28
$^{240}\mathrm{Pu} $,75.1537166934144,$7.3\cdot 10^{-1}$,\cellcolor{Gray}$4.7\cdot 10^{-1}$,1.5356891268369217,$1.7\cdot 10^{-4}$,$7.2\cdot 10^{-5}$,16.96,31.04
$^{241}\mathrm{Pu} $,13.602905556247832,$1.1\cdot 10^{-1}$,\cellcolor{Gray}$9.6\cdot 10^{-2}$,1.1667723067990616,$9.0\cdot 10^{-5}$,$6.6\cdot 10^{-5}$,71.04,90.72
$^{241}\mathrm{Am} $,0.11377797377158645,$1.7\cdot 10^{-3}$,\cellcolor{Gray}$8.0\cdot 10^{-4}$,2.181524937928079,$1.9\cdot 10^{-4}$,$4.1\cdot 10^{-5}$,100.0,100.0
\end{filecontents*}
\pgfplotstabletypeset[
		multicolumn names, 
		col sep=comma,
		columns = {Nuclide,mean_isotope,Spline-RMSE,GPR-RMSE,RatioCubicGPR,SplineRsquared,GPRRsquared,GPRPredictionin1sigma},
		columns/Nuclide/.style={
			column name=$Nuclide$, 
			string type},  
		columns/mean_isotope/.style={
			column name=$\mu$, 
			precision=2,fixed}, 
		columns/Spline-RMSE/.style={
			column name= Cubic Spline,
			string type},
		columns/GPR-RMSE/.style={
			column name=$GPR$,string type},
		columns/RatioCubicGPR/.style={
		column name=RMSE $\frac{Spline}{GPR}$,
		precision=2},
		columns/SplineRsquared/.style={
			column name=$1 - R^{2}_{Sp}$,
			string type},
		columns/GPRRsquared/.style={
			column name=$1 - R^{2}_{GPR}$,
			string type},
		columns/GPRPredictionin1sigma/.style={
			column name=$Pred-1\sigma$,
			precision=1,},
		every head row/.style={
			before row={\toprule}, 
			after row={
							&	Kg& $\mathrm{RMSE}$ & $\mathrm{RMSE}$  & &  & & $\%$ \\
				\midrule} 
		},
		every last row/.style={after row=\bottomrule}, 
		]{Sections/Tables/SobolNew169_Sobol_all_2space.csv} 
		}
			\caption{\textbf{Scenario B:} Comparison between GPR and Cubic spline models for a selection of nuclides. Highlighted cells indicate smallest error in the comparison. RMSE values are in Kg}
	\label{tab:sobolnew}
\end{table*}
\end{center}

%% file: Sections/Conclusion.tex
\section{Conclusion}\label{section:conclusion}
We have compared two interpolation methods for direct fast prediction of spent fuel nuclide compositions. As expected, GP models typically require significantly more time to interpolate values. While the time itself is of the order of miliseconds for a relatively small training set size, for large-scale applications involving a very large number of calculations for many nuclides, as is for instance the case when solving Bayesian inference problems using Markov Chain Monte Carlo \cite{ESARDA-Antonio}, it is desired to keep the required calculation time as low as possible. In return for this larger calculation time, GP models provide an estimate of the prediction uncertainty as well as an explicit method for estimating the gradient of the underlying function, provided the kernels used are differentiable. This can be useful for the solution of both forward and inverse optimization problems. 

Additionally, we have noted that GP models result in smaller interpolation errors under the $RMSE$ metric, with varying reduction factors of up to 20 when compared to Cubic Spline interpolation. 

Several possibilities exist for the extension of this work. As of now, we have created GP models depending only on two input parameters using the ASE kernel. Future research could entail the inclusion of a larger number of parameters such as power or reactor downtime, as well as the use of different kernels and even combinations of kernels for more flexible models. 

Higher fidelity GP models can be created by the use of 3-D full core reactor simulations, to account for spatial variation of nuclide concentrations and burnup, temperature and power levels. Building such GP-based models would, however, be significantly more computationally expensive. 

As we have only studied GPR to directly predict isotopic compositions, a further research area could be the study of the predictive performance when implementing GPR on a cross-section level. This could enable implementing changes in the model parameters during the irradiation cycle. This approach could make feasible the forward uncertainty estimation via Monte Carlo methods, calculating the nuclide concentrations through the matrix exponential method, all thanks to the predictive variance of the interpolated one-group cross-sections.

In conclusion, we hope that the full potential of GP-based modelling of nuclear processes will be further studied and exploited in the future, having discussed here the many advantages it has for nuclear engineering applications.

%% file: Sections/DataAvailability.tex
\section*{Data Availability}
The code used for the implementation of GP models and their comparison with Cubic Spline models can be obtained at:
\url{https://github.com/FigueroaAC/GPs-for-SpentFuel}

%% file: Sections/Acknowledgements.tex
\section*{Acknowledgments}
This research was funded by the Mathematics Division of the Center for Computational Engineering Science at RWTH Aachen University, and the Volkswagen Foundation. We would like to thank Manuel Torrilhon for his support. The funding sources had no role in the study design, collection, analysis and interpretation of data as well in the writing of the report. The calculations were run on RWTH Aachen's Compute Cluster under the rwth0572 project.

%% file: Sections/Appendix.tex
\appendix
\section{Selected nuclide models}

\begin{figure}
	\centering
	\resizebox{.95\textwidth}{!}{
		\input{Sections/Images/Sr902d}
		\input{Sections/Images/Sr901d}
	}
	\caption{Reconstruction plots for $^{90}$Sr. The left plot shows the GPR model based on samples generated by the Sobol sequence, showing the mass as a function of burnup and temperature. The reported masses have been scaled by the number of fuel assemblies in the reactor and their dimensions. The squares indicate the samples obtained from the Sobol sequence. The plot on the right shows the same model but plotting this time the predictions for two fixed temperature values alongside their posterior uncertainties. The latter are shown at the 6-$\sigma$ level. Below this plot, the errors of both curves is shown. It can be observed that they are distributed around zero.}
	\label{fig:sr90}
\end{figure}

\begin{figure}
	\centering
	\resizebox{.95\textwidth}{!}{
		\input{Sections/Images/Eu1542d}
		\input{Sections/Images/Eu1541d}
	}
	\caption{Reconstruction plots for $^{154}$Eu. The left plot shows the GPR model based on samples generated by the Sobol sequence, showing the mass as a function of burnup and temperature. The reported masses have been scaled by the number of fuel assemblies in the reactor and their dimensions. The squares indicate the samples obtained from the Sobol sequence. The plot on the right shows the same model but plotting this time the predictions for two fixed temperature values alongside their posterior uncertainties. The latter are shown at the 6-$\sigma$ level. Below this plot, the errors of both curves is shown. It can be observed that they are distributed around zero.}
	\label{fig:eu154}
\end{figure}

%% file: Sections/Images/Sr902d.tex
	\newcommand{\dsurfSr}{Sections/Images/2dSobolSr90.csv}
	\newcommand{\dsurfSrN}{Sections/Images/Sr902dComplement.csv}

	\begin{tikzpicture}
	\begin{axis}[name = surf,
		smooth,
		legend pos = north east,
		colormap/greenyellow,
		xlabel = {\textbf{Temperature} ($K$ )},
		ylabel = \textbf{Burnup} $\frac{\mathrm{MWd}}{\mathrm{KgHM}}$,
		zlabel = \textbf{Mass} (Kg),
		xlabel style={rotate=-9},
		ylabel style={rotate=35}, title=$^{90}$\textbf{Sr}
		]
		\addplot3 [surf,z buffer=sort, mesh/rows=25, opacity=0.6]
		table[y={Burnup}, x={Temperature} , z={Pred}, col sep=comma] {\dsurfSr};
		\addplot3+[only marks, scatter] table[y={Burnup}, x={Temperature} , z={Yts}, col sep=comma] {\dsurfSrN};
	\end{axis} %
\end{tikzpicture}

%% file: Sections/Images/Sr901d.tex
\newcommand{\dataSr}{Sections/Images/1dSobolSr90GPR.csv}
	\begin{tikzpicture}[scale=1]
    	\begin{axis}[name = collapsed,
    		at = (surf.south east), anchor=left of south west, xshift = 2cm,
	ylabel=\textbf{Mass} (Kg),
	grid=both,
	xtick = \empty,
	legend pos=south east, title= $^{90}$\textbf{Sr}]
	\addplot[red,only marks,mark size=1pt, mark = +] table[x={Burnup},y={Yts},col sep=comma]{\dataSr};
	\addplot[orange,thin,select coords between index={0}{24}] table [x={Burnup}, y={Pred},col sep=comma]{\dataSr};
	\addplot[green,thin,select coords between index={600}{624}] table [x={Burnup}, y={Pred},col sep=comma]{\dataSr};
	
		\addplot[name path=A,white,thin,select coords between index={0}{24}] table [x={Burnup}, 	y expr=\thisrow{Pred}+6*\thisrow{Std},col sep=comma]{\dataSr};
	\addplot[name path=B,white,thin,select coords between index={0}{24}] table [x={Burnup}, 	y expr=\thisrow{Pred}-6*\thisrow{Std},col sep=comma]{\dataSr};
	\addplot[blue!40] fill between[of=A and B];
			\addplot[name path=C,white,thin,select coords between index={600}{624}] table [x={Burnup}, 	y expr=\thisrow{Pred}+6*\thisrow{Std},col sep=comma]{\dataSr};
	\addplot[name path=D,white,thin,select coords between index={600}{624}] table [x={Burnup}, 	y expr=\thisrow{Pred}-6*\thisrow{Std},col sep=comma]{\dataSr};
	\addplot[blue!40] fill between[of=C and D];
	\legend{Test points,Prediction - 333K,Prediction - 363K,,,Uncertainty};
	
	\end{axis}

	\begin{axis}[name = errors,
	at = (collapsed.south west),
	yshift = -2cm,
	ylabel = $\bf{\Delta}$\textbf{M} (Kg),
	yscale=0.25,
	grid = both,
	xlabel =\textbf{Burnup} $\frac{\mathrm{MWd}}{\mathrm{KgHM}}$ , every axis x label/.style={at={(0.5,-3)}} ]
    \addplot[orange, select coords between index={0}{24}] table [x={Burnup}, 	y expr=\thisrow{Yts}-\thisrow{Pred},col sep=comma]{\dataSr};
	\addplot[green, select coords between index={600}{624}] table [x={Burnup}, 	y expr=\thisrow{Yts}-\thisrow{Pred},col sep=comma]{\dataSr};
    \end{axis}
	\end{tikzpicture}

%% file: Sections/Images/Eu1542d.tex
\newcommand{\dsurfEuN}{Sections/Images/Eu1542dComplement.csv}
\newcommand{\dsurfEu}{Sections/Images/2dSobolEu154.csv}
	\begin{tikzpicture}
	\begin{axis}[name = surf,
		smooth,
		legend pos = north east,
		colormap/greenyellow,
		xlabel = {\textbf{Temperature} ($K$ )},
		ylabel = \textbf{Burnup} $\frac{\mathrm{MWd}}{\mathrm{KgHM}}$,
		zlabel = \textbf{Mass} (Kg),
		xlabel style={rotate=-9},
		ylabel style={rotate=35}, title=$^{154}$\textbf{Eu}
		]
		\addplot3 [surf,z buffer=sort, mesh/rows=25, opacity=0.6]
		table[y={Burnup}, x={Temperature} , z={Pred}, col sep=comma] {\dsurfEu};
		\addplot3+[only marks, scatter] table[y={Burnup}, x={Temperature} , z={Yts}, col sep=comma] {\dsurfEuN};
	\end{axis} %
\end{tikzpicture}

%% file: Sections/Images/Eu1541d.tex
\newcommand{\dataEu}{Sections/Images/1dSobolEu154GPR.csv}
	\begin{tikzpicture}[scale=1]
    	\begin{axis}[name = collapsed,
    		at = (surf.south east), anchor=left of south west, xshift = 2cm,
	ylabel=\textbf{Mass} (Kg),
	grid=both,
	xtick = \empty,
	legend pos=north west,title=$^{154}$\textbf{Eu}]
	\addplot[red,only marks,mark size=1pt, mark = +] table[x={Burnup},y={Yts},col sep=comma]{\dataEu};
	\addplot[orange,thin,select coords between index={0}{24}] table [x={Burnup}, y={Pred},col sep=comma]{\dataEu};
	\addplot[green,thin,select coords between index={600}{624}] table [x={Burnup}, y={Pred},col sep=comma]{\dataEu};
	
		\addplot[name path=A,white,thin,select coords between index={0}{24}] table [x={Burnup}, 	y expr=\thisrow{Pred}+6*\thisrow{Std},col sep=comma]{\dataEu};
	\addplot[name path=B,white,thin,select coords between index={0}{24}] table [x={Burnup}, 	y expr=\thisrow{Pred}-6*\thisrow{Std},col sep=comma]{\dataEu};
	\addplot[blue!40] fill between[of=A and B];
			\addplot[name path=C,white,thin,select coords between index={600}{624}] table [x={Burnup}, 	y expr=\thisrow{Pred}+6*\thisrow{Std},col sep=comma]{\dataEu};
	\addplot[name path=D,white,thin,select coords between index={600}{624}] table [x={Burnup}, 	y expr=\thisrow{Pred}-6*\thisrow{Std},col sep=comma]{\dataEu};
	\addplot[blue!40] fill between[of=C and D];
	\legend{Test points,Prediction - 333K,Prediction - 363K,,,Uncertainty};
	\end{axis}

	\begin{axis}[name = errors,
	at = (collapsed.south west),
	yshift = -2cm,
	ylabel = $\bf{\Delta}$\textbf{M} (Kg),
	yscale=0.25,
	grid = both,
	xlabel =\textbf{Burnup} $\frac{\mathrm{MWd}}{\mathrm{KgHM}}$, every axis x label/.style={at={(0.5,-3)}} ]
    \addplot[orange, select coords between index={0}{24}] table [x={Burnup}, 	y expr=\thisrow{Yts}-\thisrow{Pred},col sep=comma]{\dataEu};
	\addplot[green, select coords between index={600}{624}] table [x={Burnup}, 	y expr=\thisrow{Yts}-\thisrow{Pred},col sep=comma]{\dataEu};
    \end{axis}
	\end{tikzpicture}

%% file: Paper.bbl
\begin{thebibliography}{10}
\expandafter\ifx\csname url\endcsname\relax
  \def\url#1{\texttt{#1}}\fi
\expandafter\ifx\csname urlprefix\endcsname\relax\def\urlprefix{URL }\fi
\expandafter\ifx\csname href\endcsname\relax
  \def\href#1#2{#2} \def\path#1{#1}\fi

\bibitem{nuchb}
D.~G. Cacuci, Handbook of Nuclear Engineering, Vol.~1, Springer Science \&
  Business Media, Berlin Heidelberg, 2010, Ch.~5, pp. 532--540.

\bibitem{cyclus}
K.~D. Huff, M.~J. Gidden, R.~W. Carlsen, et~al., Fundamental concepts in the
  {CYCLUS} nuclear fuel cycle simulation framework, Advances in Engineering
  Software 94 (2016) 46--59.
\newblock \href {https://doi.org/10.1016/j.advengsoft.2016.01.014}
  {\path{doi:10.1016/j.advengsoft.2016.01.014}}.

\bibitem{forensics}
K.~Dayman, S.~Biegalski, Feasibility of fuel cycle characterization using
  multiple nuclide signatures, Journal of Radioanalytical and Nuclear Chemistry
  296~(1) (2012) 195--201.
\newblock \href {https://doi.org/10.1007/s10967-012-1987-4}
  {\path{doi:10.1007/s10967-012-1987-4}}.

\bibitem{sensitivity}
E.~D. Kitcher, J.~M. Osborn, S.~S. Chirayath, Sensitivity studies on a novel
  nuclear forensics methodology for source reactor-type discrimination of
  separated weapons grade plutonium, Nuclear Engineering and Technology 51~(5)
  (2019) 1355--1364.
\newblock \href {https://doi.org/10.1016/j.net.2019.02.019}
  {\path{doi:10.1016/j.net.2019.02.019}}.

\bibitem{arp}
S.~M. Bowman, L.~C. Leal, ORIGEN-ARP: Automatic Rapid Process for Spent Fuel
  Depletion, Decay, and Source Term Analysis - NUREG/CR-0200 Revision 6, Vol.
  1-D1, Oak Ridge National Laboratory, 2006.

\bibitem{cyclus2}
A.~M. Scopatz, E.~A. Schneider, A new method for rapid computation of transient
  fuel cycle material balances, Nuclear Engineering and Design 239~(10) (2009)
  2169--2184.
\newblock \href {https://doi.org/10.1016/j.nucengdes.2009.02.022}
  {\path{doi:10.1016/j.nucengdes.2009.02.022}}.

\bibitem{baptiste}
B.~Leniau, B.~Mouginot, N.~Thiolliere, et~al., A neural network approach for
  burn-up calculation and its application to the dynamic fuel cycle code
  {CLASS}, Annals of Nuclear Energy 81 (2015) 125--133.
\newblock \href {https://doi.org/10.1016/j.anucene.2015.03.035}
  {\path{doi:10.1016/j.anucene.2015.03.035}}.

\bibitem{DOE}
L.~Diego, M.~Pedergnana, G.~Sebasti{\'{a}}n, Smart sampling and incremental
  function learning for very large high dimensional data, Neural Networks 78
  (2016) 75--87.
\newblock \href {https://doi.org/10.1016/j.neunet.2015.09.001}
  {\path{doi:10.1016/j.neunet.2015.09.001}}.

\bibitem{ESARDA-Antonio}
A.~Figueroa, M.~Goettsche, Nuclear archaeology: Reconstructing reactor
  histories from reprocessing waste, ESARDA Bulletin 59 (2019) 39--46.

\bibitem{recipes}
W.~H. Press, S.~A. Teukolsky, W.~T. Vetterling, B.~P. Flannery, Numerical
  Recipes 3rd Edition: The Art of Scientific Computing, 3rd Edition, Cambridge
  University Press, USA, 2007.

\bibitem{degradation}
P.~Baraldi, F.~Mangili, E.~Zio, A prognostics approach to nuclear component
  degradation modeling based on gaussian process regression, Progress in
  Nuclear Energy 78 (2015) 141--154.

\bibitem{BayesGP}
X.~Wu, T.~Kozlowski, H.~Meidani, K.~Shirvan, Inverse uncertainty quantification
  using the modular bayesian approach based on {G}aussian {P}rocess, part 2:
  Application to {TRACE}, Nuclear Engineering and Design 335 (2018) 417--431.

\bibitem{dmd}
R.~Elzohery, M.~Abdo, J.~Roberts, Comparison between {G}aussian {P}rocesses and
  {DMD} {S}urrogates for {I}sotopic {C}omposition {P}rediction, in: 2018 ANS
  Annual Meeting: ``Driving the Future of Nuclear Technology'', 2018.

\bibitem{na}
R.~L. Burden, J.~D. Faires, Numerical Analysis, 9th Edition, Brooks/Cole,
  Cengage Learning, 2010.

\bibitem{Candu}
W.~J. Garland (Ed.), \href{https://www.unene.ca/education/candu-textbook}{The
  Essential CANDU, A Textbook on the CANDU Nuclear Power Plant Technology},
  University Network of Excellence in Nuclear Engineering (UNENE), 2017.
\newline\urlprefix\url{https://www.unene.ca/education/candu-textbook}

\bibitem{serpent}
J.~Leppänen, M.~Pusa, T.~Viitanen, V.~Valtavirta, T.~Kaltiaisenaho, The
  {S}erpent {M}onte {C}arlo code: Status, development and applications in 2013,
  Annals of Nuclear Energy 82 (2015) 142--150.

\bibitem{sfcompo}
F.~Michel-Sendis, et~al., {SFCOMPO}-2.0: An {OECD} {NEA} {D}atabase of {S}pent
  {N}uclear {F}uel {I}sotopics assays, reactor design specifications, and
  operating data, Annals of Nuclear Energy 110 (2017) 779--788.

\bibitem{qsr}
S.~K. Sen, T.~Samanta, A.~Reese,
  \href{http://www.ijicic.org/05-036-1.pdf}{Quasi- versus pseudo-random
  generators: Discrepancy, complexity and integration-error based comparison},
  International Journal of Innovative Computing, Information and Control 2~(3)
  (2006) 621--651.
\newline\urlprefix\url{http://www.ijicic.org/05-036-1.pdf}

\bibitem{SobolSQ}
J.~Stephen, K.~Frances, Constructing {S}obol sequences with better
  two-dimensional projections, SIAM J. Sci. Comput. 30 (2008) 2635--2654.

\bibitem{NM2}
Q.~., .~Sacco, .~Saleri, Numerische Mathematik 2, Springer-Verlag, Berlin,
  2002.

\bibitem{CSplines}
C.~de~Boor, A Practical Guide to Splines, Springer-Verlag, 1978.

\bibitem{scikit-learn}
F.~Pedregosa, et~al., Scikit-learn: Machine learning in {P}ython, Journal of
  Machine Learning Research 12 (2011) 2825--2830.

\bibitem{6}
C.~E. Rasmussen, C.~K.~I. Williams, Gaussian Processes for Machine Learning,
  Massachusetts Institue of Technology, 2006.

\bibitem{scipy}
P.~Virtanen, et~al., {SciPy 1.0: Fundamental Algorithms for Scientific
  Computing in Python}, Nature Methods 17 (2020) 261--272.
\newblock \href {https://doi.org/https://doi.org/10.1038/s41592-019-0686-2}
  {\path{doi:https://doi.org/10.1038/s41592-019-0686-2}}.

\end{thebibliography}
